# Relational Lattice Foundation for Algebraic Logic

Vadim Tropashko
Vadim.Tropashko@orcl.com

___________________________________________________________________

Relational Lattice is a succinct mathematical model for Relational Algebra. It reduces the set of six classic relational algebra operators to two: natural join and inner union. In this paper we push relational lattice theory in two directions. First, we uncover a pair of complementary lattice operators, and organize the model into a bilattice of four operations and four distinguished constants. We take a notice a peculiar way bilattice symmetry is broken. Then, we give axiomatic introduction of unary negation operation and prove several laws, including double negation and De Morgan. Next we reduce the model back to two basic binary operations and twelve axioms, and exhibit a convincing argument that the resulting system is complete in model-theoretic sense. The final parts of the paper casts relational lattice perspective onto database dependency theory and into cylindric algebras.

___________________________________________________________________

## 1. INTRODUCTION

Relational Model has well established foundation in Mathematical Logic. It is often cited that relational calculus is nothing more than applied first-order predicate calculus. Can the influence be cast the other way, in other words has relational theory anything to bring onto formal logic table? In this paper we suggest that a certain offshoot of relational algebra can be viewed as a foundation for Algebraic Logic.

The simplest form of logic is propositional calculus. It has been algebraized by George Boole in the form of boolean algebra. From logical perspective sets are unary relations, so Boolean algebra is an algebra of sets. It was De Morgan who moved to the next step and established a calculus of binary relations in 1860. Charles Peirce turned out to the subject in 1870, and found most of the interesting equational laws of relation algebra. The subject fell into neglect between 1900 and 1940, to be revived by Alfred Tarski. He laid out algebraic axioms that hold in any field of binary relations hoping to find a first order characterization of fields of binary relations, the same way boolean algebra axioms characterize the field of unary relations.

The next natural step was to move into the realm of n-ary relations. The first attempt was made by Tarski with introduction of Cylindric algebras. The relation dimension moved from 2 to n, but remained fixed. This historic development prompted that the established Relation or, perhaps, Cylindric algebras were the obvious candidates for application in database management field. It turned out to be a surprise when Edgar Codd found an alternative way to generalize the logic of binary relations. Unlike Cylindric algebras, his invention -- relational algebra and relational calculus -- allowed to manipulate relations of *mixed* dimensions, and thus established the foundation of the database field.

Classic Relational Algebra employs six basic operations: projection, restriction, join, union, set difference, and renaming. These operations are not genuinely algebraic. First, the set operators – union, and difference – admit only "attribute"-compatible operands. Second, several other operations – namely, restriction, projection, and renaming – are parametrized by relation attributes. Thus, the whole system can be formally classified as a partial many-sorted algebra.

Are relational algebra operations fundamental? Darwen and Date (D&D) proposed *A New Relational Algebra* [4] (named appropriately as **A**), where the set operators were generalized to lift the aforementioned attribute compatibility limitation. In their system, join and restriction are reduced to a single logical operation ◀AND▶, while union and set difference are represented with the help of two more logical operations ◀OR▶, and ◀NOT▶. The system doesn't enjoy safety property anymore, but this is more of a problem from practical, rather than theoretical perspective. In mathematics world generalizations to infinite constructions that reduce complexity are common.

Introduction of logical operations in the spirit of boolean algebra is a welcome development. Still, in [1] we suggested that the broken absorption law is arguably the biggest technical obstacle. Moreover, the D&D system inherits "legacy" relational algebra operations of projection and renaming, which definitions appeal to relation attributes, so the whole system remains many-sorted algebra.

Relational Lattice [1-3] is an alternative representation of relational algebra in a compact system of two mathematically attractive binary operations: *natural join* and *inner union*. In this article we'll complete its axiomatization, which would be accomplished not without the help of D&D ◀OR▶ operation! The symmetry of laws of the larger system, which includes natural join, inner union, the ◀OR▶, is skewed in a remarkably peculiar way.

Relational Lattice system notation gradually evolved in the string of publications [1],[2],[3]. First, the operation symbols changed. The standard character $\bowtie$ for natural join switched to lattice theoretic $\wedge$. Likewise, the "proprietary" ⟓ glyph that we invented for inner union operation gave the way to the $\vee$. Please note that certain terminological incompatibility still remains, and we call relational operations (natural) *join* and (inner) *union* versus correspondingly lattice *meet*, and *join*. The fact that *join* term is used in both worlds to denote the exactly the opposite operation is unfortunate artifact of Relational Model legacy, which we would like to keep.

In [3] we started to extensively leverage Prover9/Mace4 [5] automated theorem proving facilities, therefore it became apparent to consolidate the notation once more. Prover9/Mace4 uses ASCII ^ (caret) for meet, and ∨ (letter *v*) for (lattice-theoretic) join. In database terminology ∧ denotes (natural) join, and ∨ – (inner) union.

The other change reflects adoption of "point-free" notation. Point-free style surfaced in database world in [8], but the ideas can be traced to functional programming, in general, and work of Roland Backhouse, in particular. Relational database theory, and predicate calculus it borrowed the foundation from, are cast in point-wise notation. They adopt function notation and put attribute names in parenthesis. For example, R(x,y) and S(y,z) are two relations named R and S with the headers consisting of the sets of attributes {x, y} and {y, z}, correspondingly. Unlike calculus, algebraic systems favor point-free notation where variable structure is deemphasized. An example analogy might be arithmetics vs. algebra. In the former we write expression whose terms refer to specific decimal structure of integer numbers, e.g.:

$$10*4 + 2 = 42$$

Arithmetics in "algebraized" form is free of any structural notation:

$$x + y = 42$$

Here the concrete appearance of symbol 42 is arithmetics legacy; it just a marker of some constant, which is considered atomic from algebraic perspective.

To summarize, in relational lattice point-free notation we operate exclusively with relation variables and constants. In rare cases, where we appeal to reader's intuition and give the definition in standard set notation, the relation symbols would be amended with the list of attributes. Point-wise definition for basic relational lattice operations is the following:

- natural join (denoted with lattice *meet* operation symbol ∧)

    $R(x,y) \wedge S(y,z) \stackrel{def}{=} \{(x,y,z) \mid (x,y) \in R \wedge (y,z) \in S\}$[1]

- inner union (denoted with lattice *join* operation symbol ∨)

    $R(y,z) \vee S(y,z) \stackrel{def}{=} \{(y) \mid \exists x\, (x,y) \in R \vee \exists z\, (y,z) \in S\}$

Apparently, at the RHS of the equality we have relations R and S with headers {x, y} and {y, z}, correspondingly. To provide the header information the LHS we have to write R(x,y) ∧ S(y,z) in point-wise notation, rather than R ∧ S in point-free form. Through the rest of the paper it would become evident that relation structure can be specified in pure algebraic constraints, that is equations and inequalities involving algebraic operations over relation variables and constants. An explicit list of relation attributes in most cases is unnecessary complication that obstructs any insight to a problem.

---

[1] This is not abuse of notation: the symbol ∧ used for lattice meet/relational join on the LHS is typographically different from logical conjunction ∧ RHS.

In addition to natural join and inner union, relational lattice lattice introduced four constants: R00 -- 0-ary empty relation, R01 -- lattice infimum, which is informally 0-ary non-empty relation, R10 -- lattice supremum, and R11 -- the "universal" relation. In [3] we argued that only R00 and R11 are genuinely interesting, and excluded the other pair from relational lattice axiom system altogether. In this paper R10 and R01 make their reappearance as essential components of bilattice axiom system. Before taking on this development, however, we have to revisit basic properties of the ◄OR► operation.

## 2. DATE&DARWEN ◄OR► OPERATION AND ITS DUAL

The ◄OR► operation can be defined either in set notation (point-wise)

$$R(x,y) \blacktriangleleft OR \blacktriangleright S(y,z) \stackrel{def}{=} \{(x,y,z) \mid (x,y) \in R \land z \in Z \lor x \in X \land (y,z) \in S \}$$

or, alternatively, in the relational lattice terms (point-free)

$$R \blacktriangleleft OR \blacktriangleright S \stackrel{def}{=} (R \wedge (S \vee R11)) \vee (S \wedge (R \vee R11)).$$

where R11 is the universal relation [2,3]. In order to continue leveraging Prover9 facilities we found it convenient to switch the notation to the plus symbol +. Then, we proved associativity of the + and distributivity of natural join ∧ over the +

$$Q + (R + S) = (Q + R) + S.$$
$$Q \wedge (R + S) = (Q \wedge R) + (Q \wedge S).$$

in the relational lattice system. To round up this development let's call the + operation as *outer union*.

Let introduce a new operation * -- *inner join*, which is dual to the outer union. In set notation

$$R(x,y) * S(y,z) \stackrel{def}{=} \{(y) \mid \exists x\, (x,y) \in R \land \exists z\, (y,z) \in S \}$$

while in the relational lattice terms

$$R * S \stackrel{def}{=} (R \vee (S \wedge R00)) \wedge (S \vee (R \wedge R00)).$$

This was the last occurrence where set notation was used. From now on we switch to lower letters relation names and write x and y instead of R and S.

The following theorem, which claims absorption property for outer union and inner join, proved to be critical for further development

$$x + (x * y) = x.$$

These results prompt that inner join and outer union operations form another lattice structure. Unfortunately, this conjecture is wrong. Inner join is not associative

~~(x * y) * z = x * (y * z).~~

Next, absorption of inner join and outer union is not a valid proposition, either

~~$x + (x * y) = x.$~~

These two negative results were established with one more computer system – QBQL [6]. Its primary purpose is to be able to check if an assertion is a valid proposition in relational lattice. Compared to Mace4, QBQL invalidates wrong assertions by finding counterexample relation objects, whereas Mace4 works with generic models where algebraic elements assumed to be atomic.

If the inner join and outer union formed a lattice, what were the top and bottom elements? Here are two more theorems

$x * R00 = R00.$

$x * R11 = x.$

Then, "dually"

$x + R00 = x.$

$x + R11 = R11.$

In other words, the R00 and R11 elements in (broken) inner join/natural union lattice play the roles of the R01 and R10 and vice versa. These new results force the reevaluation of the entire relational lattice axiom system.

3. BILATTICE AXIOM SYSTEM

Now that we established the secondary lattice structure, it is natural to redefine bi-lattice structure explicitly rather than derive it from the single lattice axiom system. We have bi-lattice of four operations and four constants

$x \wedge y = y \wedge x.$
$(x \wedge y) \wedge z = x \wedge (y \wedge z).$
$x \wedge (x \vee y) = x.$
$x \vee y = y \vee x.$
$(x \vee y) \vee z = x \vee (y \vee z).$
$x \vee (x \wedge y) = x.$

$x * x = x.$
$x * y = y * x.$
~~$(x * y) * z = x * (y * z).$~~
$x * (x + y) = x.$
$x + x = x.$
$x + y = y + x.$
$(x + y) + z = x + (y + z).$
~~$x + (x * y) = x.$~~

$x \wedge R10 = R10.$
$x \vee R01 = R01.$
$x * R00 = R00.$
$x + R11 = R11.$

Since the * + lattice laws are (partially) broken, we have to explicitly add idempotent laws. The identities

**x * y = (x v (y ^ R00)) ^ (y v (x ^ R00)).**
**x + y = (x ^ (y v R11)) v (y ^ (x v R11)).**
**x ^ y = (x + (y * R10)) * (y + (x * R10)).**
**x v y = (x * (y + R01)) + (y * (x + R01)).**

provide explicit "definitions" connections one pair of operations in terms of the other.

In [3] we have established the distributivity of inner join over outer union. It is natural to wonder if inner union distributes over inner join, and whether the picture is partially symmetric, at least. The answer is to this question is negative. Again, QBQL [6] is our standard tool when it comes to establishing negative propositions in relational lattice. Two final laws added to the system:

**x ^ (y + z) = (x ^ y) + (x ^ z).**
**x + (y ^ z) = (x + y) ^ (x + z).**

Not all the axioms are independent: the chosen presentation appeals to the symmetry of the system. Unlike bi-lattices known in the literature ([7]), our secondary lattice has some laws broken (which formally disqualifies the whole system as "bi-lattice"). Fig. 1 summarizes the connection between various pairs of operations.

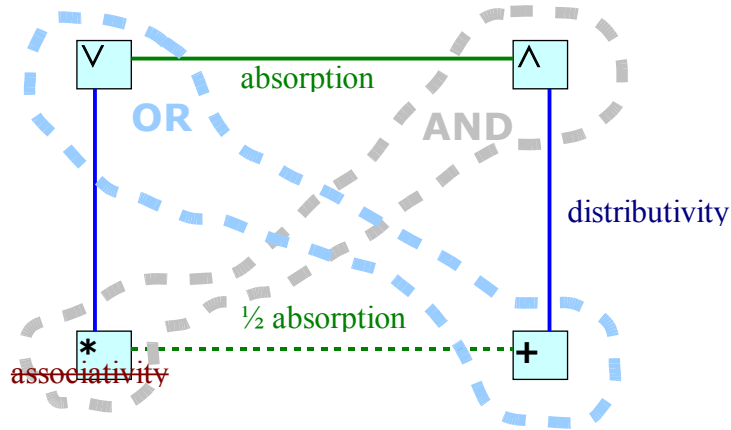

Figure 1. Connections among various pairs of bi-lattice operations. Natural join and inner join generalize the AND operation from boolean algebra, while outer union and inner union extend the OR.

In our new system Fundamental Decomposition Identity (FDI, [2,3]) and its cousin over inner join and outer union are both theorems:

**x = (x ^ R00) v (x ^ R11).**
**x = (x + R01) * (x + R10).**

Weaker forms of dual to FDI law (which was called FDI$^{-1}$ in [3]) together with its cousin are theorems as well:

    R00 ^ (x v R11) = x ^ R00.
    R01 + (x * R10) = x + R01.

"Alternative definition" for outer union is one more theorem

    x + y = (x ^ (y v R11)) v (y ^ (x v R11)).

4. COMPLEMENT

In [3] we have introduced binary anti-join operation as generalization of set difference. It can be expressed in terms of unary operation that is essentially equivalent to D&D ◄NOT►. An unary complement operator, which in compliance with Prover9 we denote by ASCII single quote ', is defined by the following pair of axioms:

    **x' ^ x = x ^ R00.**
    **x' v x = x v R11.**

Double negation

    x'' = x.

and De Morgan

    x' + y' = (x ^ y)'.

are both theorems in the bi-lattice system. Other theorems involving complement include:

    (x + R01)' = x ^ R00.
    (x * R10)' = x v R11.
    (x ^ R00)' = x v R11.
    (x + R01)' = x * R10.
    R11' = R10.
    R00' = R01.

5. NAND BASIS

The following three theorems are also provable automatically in bi-lattice axiom system:

    x ^ y = ((x ^ y)' ^ (x ^ y)')'.
    x + y = ((x ^ x)' ^ (y ^ y)')'.
    x' = (x ^ x)'.

They establish that binary NAND operation can be viewed as a basis for a fragment of D&D algebra that involves logical operations ◄AND►, ◄OR►, and ◄NOT►. The significance of this reduction is unclear: once again the pair of ◄AND► and ◄OR► operations is not a lattice, let alone boolean algebra. Therefore, the utility of NAND in boolean algebra might not necessarily carry over into relational algebra.

6. MINIMALISTIC AXIOMATIZATION

Let's turn to previously ignored issues of independence and completeness. Are really that many operations and axioms required? What about other axioms, such as Spight Distributivity Criteria (SDC) [3]; do they all become theorems in the new system? Reducing the system may help answering these questions. Such reduction may also benefit our leverage of Prover9 system, because extra operations and extra axioms explode the search space.

First, the theorems

```
R10 = R11'.
R01 = R00'.
```
or, alternatively
```
R10 = R11 ^ R00.
R01 = R11 v R00.
```
render the constants R10 and R01 redundant. Next, the defining axioms for inner join and outer union
```
x * y = (x v (y ^ R00)) ^ (y v (x ^ R00)).
x + y = (x ^ (y v R11)) v (y ^ (x v R11)).
```
can be used to replace these operations in terms of natural join and inner union. The resulting system features two constants (that is 0-ary operations), one unary operation, two binary ones, and twelve axioms:

```
x ^ y = y ^ x.
(x ^ y) ^ z = x ^ (y ^ z).
x ^ (x v y) = x.
x v y = y v x.
(x v y) v z = x v (y v z).
x v (x ^ y) = x.

x = (x ^ R00) v (x ^ R11).
(x v (y ^ R00)) ^ (y v (x ^ R00))=(x ^ y) v ((x v y) ^ R00).
x ^ ((y' ^ z'))' = ((x ^ y)' ^ (x ^ z)')'.
R00 ^ (x ^ (y v z)) = R00 ^ ((x ^ y) v (x ^ z)).
x' ^ x = x ^ R00.
x' v x = x v R11.
```

The lattice part of the system is ubiquitous, and therefore warrants no comment. The FDI axiom [3] being excluded from bi-lattice system make its reappearance. The second on the list of non-lattice theoretic axioms is an identity which LHS and RHS represent the two alternative definitions for inner join. The third axiom is distributivity of natural join over outer union. The fourth axiom is Distributivity Constraint on relation Headers (DCH, [3]). The last two axioms define complement operation.

Let's verify the consistency of the proposed system. The natural test would be to input general distributivity identity

$$(x \wedge (y \vee z) = (x \wedge y) \vee (x \wedge z)).$$

into Mace4 and witness it producing 6-element counterexample

```
^ :
      | 0 1 2 3 4 5
    ---+------------
    0 | 0 4 4 0 4 4
    1 | 4 1 2 1 4 5
    2 | 4 2 2 2 4 4
    3 | 0 1 2 3 4 5
    4 | 4 4 4 4 4 4
    5 | 4 5 4 5 4 5

v :
      | 0 1 2 3 4 5
    ---+------------
    0 | 0 3 3 3 0 3
    1 | 3 1 1 3 1 1
    2 | 3 1 2 3 2 1
    3 | 3 3 3 3 3 3
    4 | 0 1 2 3 4 5
    5 | 3 1 1 3 5 5

' :
        0 1 2 3 4 5
    ----------------
        3 4 5 0 1 2

    R00 : 0
    R11 : 1
```

After all, if no counterexample were found the suspicion might be that the whole system collapses to boolean algebra (thus, rendering the whole paper as vacuous)!

Mace4 considers the model elements atomic, they are labeled by integers from 0 to 5. Can these elements be interpreted as relations, so that operations of inner union, natural join, and complement defined in point-wise terms are consistent with the above? A reader can verify that the following mapping works:

```
    R00                <--> 0
    R01  (=R00')       <--> 3
    {(t=a)}            <--> 5
    {(t=b)}            <--> 2
    {(t=a),(t=b)}      <--> 1
    empty(t)           <--> 4
```

The mapping preserves relational operations, for example, a point-wise equality

$$\{(t=a)\} \wedge \{(t=b)\} = \text{empty}(t)$$

corresponds to point-free identity

```
        5 ^ 2 = 4
```

read off the ^ multiplication table.

Such model search can also cast some light into the following question: How many relations with empty attribute sets exist? We certainly know two: `R00` and `R01`, but what if there is more? This question translates directly into the goal

```
        x v R00 = R00 | x v R00 = R00'.
```

With this input Mace4 produces 4 element model, such that `R11 < R00`. This might considered a legitimate relational lattice, but for our purposes lets postulate that `R11` and `R00` are incompatible in the lattice order:

```
        R00 ^ R11 != R00.
```

With this amendment Mace4 switches to 8 element model:

```
  ^ :
         | 0 1 2 3 4 5 6 7
      ---+----------------
       0 | 0 3 0 3 0 3 0 3
       1 | 3 1 5 3 1 5 7 7
       2 | 0 5 2 3 2 5 0 3
       3 | 3 3 3 3 3 3 3 3
       4 | 0 1 2 3 4 5 6 7
       5 | 3 5 5 3 5 5 3 3
       6 | 0 7 0 3 6 3 6 7
       7 | 3 7 3 3 7 3 7 7

  v :
         | 0 1 2 3 4 5 6 7
      ---+----------------
       0 | 0 4 2 0 4 2 6 6
       1 | 4 1 4 1 4 1 4 1
       2 | 2 4 2 2 4 2 4 4
       3 | 0 1 2 3 4 5 6 7
       4 | 4 4 4 4 4 4 4 4
       5 | 2 1 2 5 4 5 4 1
       6 | 6 4 4 6 4 4 6 6
       7 | 6 1 4 7 4 1 6 7

  ' :
           0 1 2 3 4 5 6 7
         ------------------
           4 3 6 1 0 7 2 5

R00 : 0
R11 : 1
c1 : 2
```

Is there a point-wise interpretation of the above result? Here is one possibility:

```
empty(t)                  <-->  0
{(t=a)}                   <-->  2
{(t=b)}                   <-->  6
{(t=a),(t=b)}             <-->  4
empty(t,s)                <-->  3
{(t=a,s=1)}               <-->  5
{(t=b,s=2)}               <-->  7
{(t=a,s=1),(t=b,s=2)}     <-->  1
```

This mapping seems to make no sense: how can relation R00 with empty set of attributes map into unary relation `empty(t)`? However, there is nothing in the point-free system that can possibly convey this restriction. According to relational lattice axioms R00 is some relation constant, which among all other lattice elements is the best candidate onto the role of empty relation with empty set of attributes. On "poor relation instance market" the system has to choose a candidate with nonempty set of attributes.

Database practice suggests another perspective onto this surprising result. All relations in the system can be equipped with a "hidden" column (`ROWID` in oracle). There is one subtle but significant detail, however: in relational lattice natural join can't just project away `ROWID` columns as oracle does, otherwise the system would be inconsistent. Consider the alternative – joining by `ROWID` values – which is never meaningful.

## 7. CONDITIONAL DISTRIBUTIVITY

The theme of broken distributivity law originated in [1], is traced through subsequent publications [2,3]. In particular, [2] suggested that "*Evidently, distributivity holds when there is something special about relation headers (e.g all relations have the same header), or there is something special about relation content (e.g. all relations are empty)*". This section would clarify this claim.

The system from section 6 lists two distributivity related axioms, while SDC

$$\mathbb{R}00 \wedge (x \vee y) = \mathbb{R}00 \wedge (x \vee z) \rightarrow x \wedge (y \vee z) = (x \wedge y) \vee (x \wedge z).$$

is not included. Is it a theorem? Several attempts to establish it (with the help of Prover9) failed. The model search confirmed its validity up to 24 elements domain size. Proving weaker criteria

$$R00 \wedge y = R00 \wedge z \rightarrow x \wedge (y \vee z) = (x \wedge y) \vee (x \wedge z).$$

takes several hours of Prover9 run. One more conditional distributivity theorem

$$R11 \wedge y = R11 \wedge z \rightarrow x \wedge (y \vee z) = (x \wedge y) \vee (x \wedge z).$$

supports the earlier intuition that distributivity holds when relations y and z have the same content.

These partially incomplete results seem to support the idea that various conditional distributivity identities (including SDC), are just reflections of universal distributivity of join over outer union. The significance of the other axiom – DCH – is not clear.

## 8. LITTLE DEPENDENCY THEORY

Dependency theory is considered one of gems of the database field. Again, has relational lattice theory anything to offer? This section uses the standard definition of lattice order

> x < y <-> x ^ y = y.

It would be demonstrated that database constraints are algebraic order constraints.

Let's warm up on inclusion dependencies. Assume the relations r and s have the same set of attributes x. In relational lattice a set of attributes is abstracted as an empty relation. By abuse of notation let's keep the same variable name x. Formally,

> R00 ^ x = x.
> (r v s) ^ R00 = x.

The inclusion dependency (also known as referential integrity constraint) asserts that projection of s onto the set of attributes x is a subset of r projected over x. In relational lattice terms this is formally expressed as

> r v x < s v x.

It is natural to suggest that the former two identities are inessential to the problem at hand. The *generalized inclusion dependency* is defined to satisfy the above order constraint regardless of the other restrictions upon r, s and x. In particular, x doesn't have to be empty. Inclusion dependency is transitive, and indeed establishing

> r v x < s v x &
> s v y < s v y &
> x < y ->
> r v x < s v y.

is an easy Prover9 exercise. The assertion x < y generalizes the condition that the set of attributes x must be a subset of y.

Let's turn to functional dependency. This turned to be a much tougher subject. First, the standard definition of functional dependency needs an equality relation, and relational lattice doesn't have one! The other technical difficulty is renaming operation, that we managed to avoid in previous development. Renaming can be expressed with the help of equality ([1]), but regression to point-wise constructions seems to be unavoidable.

One approach is to lower abstraction level, and consider *single tuple* functional dependency. The statement "Every time Max is at Home, Claire is at the Library"[2] is an example of such dependency. Let's make few observations. First, logic perspective becomes apparent. At least, we can progress and formulate the above statement in terms of relational lattice operations. It is still not clear how to elevate *single tuple* dependency

---

[2]Borrowed from "*Language Proof and Logic*" textbook by Jon Barwise and John Etchemendy

to functional dependency. Quantification is a point-wise construction that arguably has no place in the relational lattice system. Let's ignore this issue for now and focus on "tuple" part of the concept. In point-free system, again, we don't operate tuples. Therefore, lets just generalize tuples into relations. What is left of the former dependency (which we can't even call *single tuple* anymore) is this constraint:

$$(r \wedge x) \wedge y < (r \wedge x) \wedge y'.$$

where the is the original relation[3] r and two other relations, the function argument x and return value y. At this point it looks like the analogy to ordinary functional dependency is too far stretched, and our construction is nothing of resemblance. Let's not give up, however, and introduce new relation symbol[4]

$$FD(r,x,y) <-> (r \wedge x) \wedge y < (r \wedge x) \wedge y'.$$

"FD" stands for "fictional dependency", and point-wise notation is unavoidable, because it is a relation over relations. Here are theorems

```
y < x -> FD(r,x,y).
FD(r,x,y) & FD(r,y,z) -> FD(r,x,z).
FD(r,x,y) -> FD(r,x^z,y^z).
```

Modulo notation, those are Armstrong axioms! The first two propositions surrender to Prover9 brute force; reflexivity is immediate, while transitivity takes less than 2 hours run time. The augmentation is proved in Appendix A.

9. CYLINDRIFICATION

*Cylindric set algebra* [9] employs unary operator $c_K$, which transforms every n-ary relation into (n-1)-ary one by "forgetting" the k-th argument. Relational lattice analog is binary *cylindrification* operation defined as follows

$$y @ x \stackrel{\text{def}}{=} (y \vee R11) + x.$$

Its properties

```
x @ x = x v R11.
(x @ x) @ x = x @ x.
z @ (x ^ (z @ y)) = z @ (x ^ (z @ y)).
x @ y = y @ x.
z @ (x + y) = (z @ x) + (z @ y).
z @ (x ^ y) = (z @ x) ^ (z @ y).
z @ (x v y) = (z @ x) v (z @ y).
x @ (y @ z) = (x @ y) @ z.
```

follow from relational lattice axioms.

---

[3] that is "instance" in dependency theory terminology

[4] Here the term *relation* become overloaded. So far we dealt with operations upon relations, but now a relation over relations is introduced. Perhaps, introducing a sloppy term "relationship", would eliminate source of confusion. In that terminology x < y is the order *relationship*.

## 10. CONCLUSION

Database theory accumulated volumes of research, and it is still unclear how much of it can be cast into algebraic perspective. In earlier work we provided a convincing argument that query transformation part of modern optimizers can be based upon relational lattice system. In this paper we scratched the surface of dependency theory, although most of the paper is focused on more ambitious claim of logic algebraizaition. Certainly, relational algebra is as expressive as relational calculus, which in turn is nothing more than applied predicate logic. As it has been demonstrated, cylindric algebra is expressible in our system, and its properties follow from relational lattice axioms. Yet, it's unclear how relational lattice matches against another recognized system -- binary relation algebra. On one hand, relational lattice elements are more general: they are relations of arbitrary mixed dimensions. However, binary relation algebras include equality relation and capture some important operations such as transitive closure, both missing in the relational lattice framework.

## 11. ACKNOWLEDGMENTS

I'm indebted to Michael K. Kinyon who provided several hints on Prover9 help forum.

## APPENDIX A. AUGMENTATION AXIOM PROOF

Prover9-assisted manual proof of the augmentation axiom. Goal:

```
(r^x)^y < (r^x)^y' -> (r^(x^z))^(y^z) < (r^(x^z))^(y^z)'.
```

Simplify implication RHS:

```
(r^x)^y < (r^x)^y' -> (r^(x^z))^y < (r^(x^z))^(y^z)'
```

Use De Morgan law (expressed in terms of inner union and natural join)

```
(y^z)' = (z' ^ (y' v R11)) v (y' ^ (z' v R11)).
```

to expand implication RHS. Check [weak] conditional distributivity:

```
R00 ^ (z' ^ (y' v R11)) = R00 ^ (y' ^ (z' v R11)).
```

The expanded implication to prove:

```
(r^x)^y < (r^x)^y' ->
(r^(x^z))^y < ( (r^(x^z))^(z'^(y' v R11)) ) v ( (r^(x^z))^(y'^(z' v R11)) ).
```

Split it into two individual proofs:

```
(r^x)^y < (r^x)^y' -> (r^(x^z))^y < (r^(x^z))^(z'^(y' v R11)).
```

proved automatically, and

```
(r^x)^y < (r^x)^y' -> (r^(x^z))^y < (r^(x^z))^(y'^(z' v R11)).
```

Split the last implication into the two:

```
(r^x)^y < (r^x)^y' -> (r^(x^z))^y < (r^(x^z))^y'.
```

```
(r^(x^z))^y < (r^(x^z))^y' -> (r^(x^z))^y < (r^(x^z))^(y'^(z' v R11)).
```

The second is immediate. The a substitution reduces the first one to

```
t^y < t^y' -> (t^z)^y < (t^z)^y'.
```

which is immediate as well.